\documentclass[10pt, onecolumn, conference]{IEEEtran}

\usepackage{graphicx,url}
\usepackage{tabularx}
\usepackage{array}
\usepackage[utf8]{inputenc}  
\usepackage{hyperref}
\hypersetup{
    unicode=true,
    breaklinks=true,
    colorlinks=true,
    linkcolor=black,
    citecolor=black,
    urlcolor=black
}
     
\title{Temperature in SLMs: Impact on Incident Categorization in On-Premises Environments}

\author{
\IEEEauthorblockN{
    Marcio Pohlmann\IEEEauthorrefmark{1},
    Alex Severo\IEEEauthorrefmark{1},
    Gefté Almeida\IEEEauthorrefmark{1} \\
    Diego Kreutz\IEEEauthorrefmark{1},
    Tiago Heinrich\IEEEauthorrefmark{2},
    Lourenço Pereira\IEEEauthorrefmark{3}
}
\IEEEauthorblockA{\IEEEauthorrefmark{1}AI Horizon Labs, Federal University of Pampa (UNIPAMPA)}
\IEEEauthorblockA{\IEEEauthorrefmark{2}Max Planck Institute for Informatics (MPI)}
\IEEEauthorblockA{\IEEEauthorrefmark{3}Instituto Tecnológico de Aeronáutica (ITA)}
}

\begin{document} 

\maketitle

\begin{abstract}

SOCs and CSIRTs face increasing pressure to automate incident categorization, yet the use of cloud-based LLMs introduces costs, latency, and confidentiality risks. We investigate whether locally executed SLMs can meet this challenge. We evaluated 21 models ranging from 1B to 20B parameters, varying the temperature hyperparameter and measuring execution time and precision across two distinct architectures. The results indicate that temperature has little influence on performance, whereas the number of parameters and GPU capacity are decisive factors.

\end{abstract} 


\section{Introduction}

The increasing volume and complexity of cybersecurity incidents have generated a growing overload on response teams, which face the need for scalable solutions for triage, categorization, and prioritization of events. Structured categorization of these incidents is essential for identifying patterns, understanding the diversity of threats, and improving defense strategies. However, this process still faces important limitations, such as ambiguity in reports, a lack of standardization, and a shortage of specialized professionals. 

In this scenario, automated approaches based on artificial intelligence emerge as promising alternatives to accelerate categorization and increase operational efficiency \cite{nplcyberincsec:2024}. Despite this, significant challenges persist, associated with the lack of labeled data, semantic ambiguity, and variability in attack formats \cite{Ibrishimova2019Cyber}. Recent studies indicate that hybrid approaches combining human expertise and advanced computational techniques have the potential to mitigate these limitations and make incident classification more robust, accurate, and aligned with real decision-making needs. 
Among these approaches, Large Language Models (LLMs) stand out for their ability to process unstructured text, establish complex semantic correlations, and generate consistent inferences with high efficiency.

Conversely, the costs associated with LLM usage, as well as the need to anonymize sensitive incident-related information, may limit the adoption of these tools in corporate environments. In this context, Small Language Models (SLMs) gain relevance, as they can be executed locally (on-premises) and require less computational capacity. Moreover, these models allow fine-tuning through the selection of architectures with different numbers of parameters and calibration of inference hyperparameters, such as sampling (\textit{top-k}, \textit{top-p}, and \textit{temperature}), penalization (\textit{frequency}, \textit{presence}, and \textit{repetition}), and generation control (\textit{max\_tokens}, \textit{min\_tokens}, and \textit{stop}) \cite{zhao2023survey}.

In this work, we investigate the influence of the temperature hyperparameter in SLMs from different vendors and of different magnitudes, evaluating both execution time and precision in incident categorization. For the experiments, we used a dataset composed of six balanced categories of real incidents from a CSIRT, enabling a consistent assessment of the models’ classification capabilities. The main contribution of this study is a systematic and comprehensive experimental evaluation involving 21 distinct models, offering a comparative analysis of the impact of temperature on SLMs executed locally for security incident categorization.


\section{Parameters and Hyperparameters in Language Models}

In language models based on deep neural networks, parameters correspond to the values learned during training, such as weights and biases that transform input signals across layers\footnote{\url{https://www.ibm.com/think/topics/model-parameters.}}
. The number of parameters determines the model’s representational capacity: larger architectures capture more complex linguistic relationships but require more computational resources and present a higher risk of \textit{overfitting}.

Hyperparameters are defined before training or during inference, and they influence both the learning process and the model’s behavior. In the context of inference, these hyperparameters can be organized into three categories: sampling, penalization, and generation control. Sampling hyperparameters determine how the next token is selected, including temperature, \textit{top-k}, and \textit{top-p}. Penalization hyperparameters adjust probabilities to avoid excessive repetition, while control hyperparameters define structural limits of the output, such as the maximum number of tokens and \textit{stop sequences} \cite{zhao2023survey}.

Temperature regulates the level of randomness in text generation. Low values tend to produce more deterministic and precise outputs \cite{Renze_2024}, while higher values increase diversity and the risk of incoherence \cite{wang2020contextualtemperaturelanguagemodeling}. Studies indicate that temperatures close to 0.0 are suitable for reasoning and translation, temperatures above 1.0 favor creativity, and values higher than 1.6 may lead to the so-called temperature paradox \cite{Renze_2024, li2025exploringimpacttemperaturelarge}.

\section{Methodology}\label{metodologia}

\begin{figure}[!ht]
    \centering
    \includegraphics[width=1\linewidth]{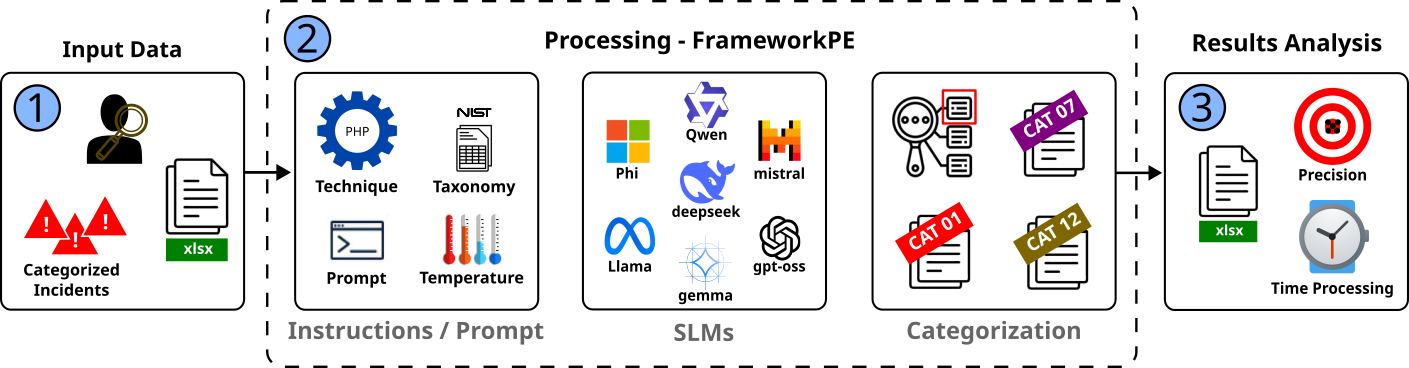}
    \caption{Experiment Stages}
    \label{fig:experimento}
\end{figure}

The experiments were structured as illustrated in \autoref{fig:experimento}, following a three-stage pipeline similar to that adopted in recent studies (e.g., \cite{severo_sbseg_estendido,severo_sbseg}): Input Data, Processing, and Results Analysis. The objective was to evaluate the automated categorization of security incidents considering variations in temperature, processing time, and precision. For the experiments, two distinct computational architectures were used: (i) an AMD Ryzen 7 4800H with 32 GB of RAM and an NVIDIA GeForce GTX 1650 GPU (4 GB), and (ii) an Intel Core i7-12700 with 64 GB of RAM and an NVIDIA RTX A4000 GPU (16 GB).

Regarding the \textbf{Input Data} used to evaluate the performance of SLMs in classifying security incidents, we employed the dataset of real CSIRT incidents created and described by \cite{severo_sbseg_estendido}. The original dataset contains 194 anonymized records, from which we selected a balanced subset composed of six distinct categories, each containing four incidents. The number of incidents per class was determined by the minority classes. The entire original dataset had previously been classified by two cybersecurity specialists, who independently analyzed the incidents, establishing a ground truth for comparison with the automated results.

In the \textbf{Processing} stage, we used FrameworkPE\footnote{\url{https://github.com/AILabs4All/FrameworkPE}}, which implements several prompt engineering techniques. The user can select the desired technique, and for our experiments, we adopted PHP, which achieved the best results among the techniques recently evaluated for security incident classification \cite{severo_sbseg}. PHP is complemented by the use of a baseline taxonomy (NIST), textual rules for output standardization, parametrization of inference temperature, and the selection of different SLMs.

We included models ranging from 1 to 20 billion parameters made available by the Ollama provider (version 0.12.3), totaling 21 language models from seven different vendors. Each execution produces a collection of categorized incidents, which is subsequently used in the evaluation stage.
The final stage, \textbf{Results Analysis}, consisted of measuring precision (comparison between inference and the ground truth) and processing time obtained by each model, considering the two hardware architectures used in the evaluation: Ryzen7/GTX 1650 and i7/RTX A4000.

\section{Results and Discussion}
\label{resultados}

Table \ref{tab:execution_times} presents the execution times of the 21 evaluated language models under four temperature configurations (T0, T0.4, T0.7, and T1). The values are displayed in the H:MM:SS format, including totals per model and per temperature configuration.

\begin{table}[!htp]
    \centering
    \caption{Execution times of the models under different temperatures}
    \label{tab:execution_times}
    \renewcommand{\arraystretch}{1.3}
    \resizebox{\textwidth}{!}{%
    \begin{tabular}{l|cccc|c|cccc|c}
    \hline
    & \multicolumn{5}{c|}{\textbf{Ryzen 7 / GTX 1650 Architecture}} & \multicolumn{5}{c}{\textbf{i7/RTX A4000 Architecture}} \\
    \hline
    \textbf{Model} & \textbf{T0} & \textbf{T0.4} & \textbf{T0.7} & \textbf{T1} & \textbf{Model Total} & \textbf{T0} & \textbf{T0.4} & \textbf{T0.7} & \textbf{T1} & \textbf{Model Total} \\
    \hline
    gemma3:1b & 0:04:57 & 0:04:38 & 0:04:31 & 0:04:30 & 0:18:36 & 0:01:17 & 0:01:23 & 0:01:54 & 0:02:27 & 0:07:01 \\
    gemma3:4b & 0:14:08 & 0:13:59 & 0:13:56 & 0:14:21 & 0:56:24 & 0:01:48 & 0:05:20 & 0:05:15 & 0:02:32 & 0:14:55 \\
    gemma3:12b & 1:26:06 & 1:24:42 & 1:24:58 & 1:25:12 & 5:40:58 & 0:03:47 & 0:06:21 & 0:10:36 & 0:24:06 & 0:44:50 \\
    phi4:14b & 1:03:02 & 1:05:13 & 0:59:10 & 1:01:52 & 4:09:17 & 0:10:01 & 0:09:38 & 0:18:19 & 0:15:12 & 0:53:10 \\
    phi3:3.8b & 0:23:11 & 0:24:39 & 0:21:40 & 0:22:08 & 1:31:38 & 0:03:08 & 0:07:09 & 0:15:28 & 0:04:16 & 0:30:01 \\
    phi3:14b & 1:07:18 & 1:03:48 & 1:04:16 & 1:06:37 & 4:21:59 & 0:04:18 & 0:04:26 & 0:10:32 & 0:24:36 & 0:43:52 \\
    deepseek-r1:1.5b & 0:19:28 & 0:17:36 & 0:19:06 & 0:18:09 & 1:14:19 & 0:11:29 & 0:10:04 & 0:14:28 & 0:07:47 & 0:43:48 \\
    deepseek-r1:7b & 1:18:48 & 1:22:36 & 1:22:30 & 1:29:46 & 5:33:40 & 0:24:23 & 0:14:21 & 0:24:48 & 0:15:51 & 1:19:23 \\
    deepseek-r1:8b & 2:30:37 & 2:25:35 & 2:26:27 & 2:25:33 & 9:48:12 & 0:35:46 & 0:37:30 & 0:30:38 & 0:23:06 & 2:07:00 \\
    deepseek-r1:14b & 1:46:26 & 1:46:29 & 1:44:51 & 1:45:39 & 7:03:25 & 0:12:19 & 0:29:25 & 0:24:09 & 0:24:36 & 1:30:29 \\
    qwen3:1.7b & 0:16:37 & 0:15:09 & 0:15:29 & 0:15:39 & 1:02:54 & 0:10:12 & 0:12:01 & 0:11:16 & 0:12:37 & 0:46:06 \\
    qwen3:4b & 1:26:25 & 1:30:29 & 1:25:05 & 1:23:59 & 5:45:58 & 0:33:41 & 0:33:21 & 0:29:10 & 0:43:03 & 2:19:15 \\
    qwen3:8b & 1:35:15 & 1:32:21 & 1:33:43 & 1:33:43 & 6:15:02 & 0:23:59 & 0:12:12 & 0:22:03 & 0:18:47 & 1:17:01 \\
    qwen3:14b & 3:01:25 & 2:57:36 & 3:02:34 & 3:01:16 & 12:02:51 & 0:18:15 & 0:31:35 & 0:28:35 & 0:24:58 & 1:43:23 \\
    llama3.2:1b & 0:15:13 & 0:13:48 & 0:11:29 & 0:14:19 & 0:54:49 & 0:06:25 & 0:15:24 & 0:10:34 & 0:13:20 & 0:45:43 \\
    llama3.2:3b & 0:15:01 & 0:15:38 & 0:17:09 & 0:15:31 & 1:03:19 & 0:07:17 & 0:05:37 & 0:09:29 & 0:07:57 & 0:30:20 \\
    llama3.1:8b & 0:25:43 & 0:31:28 & 0:33:44 & 0:35:06 & 2:06:01 & 0:12:14 & 0:15:14 & 0:23:42 & 0:12:20 & 1:03:30 \\
    llama3:8b & 0:25:43 & 0:25:23 & 0:25:48 & 0:25:42 & 1:42:36 & 0:23:00 & 0:04:51 & 0:11:40 & 0:06:48 & 0:46:19 \\
    mistral:7b & 0:32:04 & 0:30:34 & 0:31:21 & 0:31:30 & 2:05:29 & 0:04:28 & 0:04:41 & 0:03:41 & 0:05:08 & 0:17:58 \\
    mistral-nemo:12b & 0:43:55 & 0:43:14 & 0:43:51 & 0:42:55 & 2:53:55 & 0:05:58 & 0:13:31 & 0:04:03 & 0:05:33 & 0:29:05 \\
    gpt-oss:20b & 1:03:14 & 1:05:48 & 1:03:40 & 1:01:25 & 4:14:07 & 0:04:01 & 0:03:44 & 0:09:15 & 0:15:00 & 0:32:00 \\
    \hline
    \textbf{Temperature Total} & \textbf{20:14:36} & \textbf{20:10:43} & \textbf{20:05:18} & \textbf{20:14:52} & & \textbf{4:17:46} & \textbf{4:37:48} & \textbf{5:19:35} & \textbf{5:10:00} & \\
    \textbf{Architecture Total} & & & & & \textbf{80:45:29} & & & & & \textbf{19:25:09} \\
    \hline
    \end{tabular}%
    }
\end{table}

\begin{figure}[h]
    \centering
    \includegraphics[width=1\linewidth]{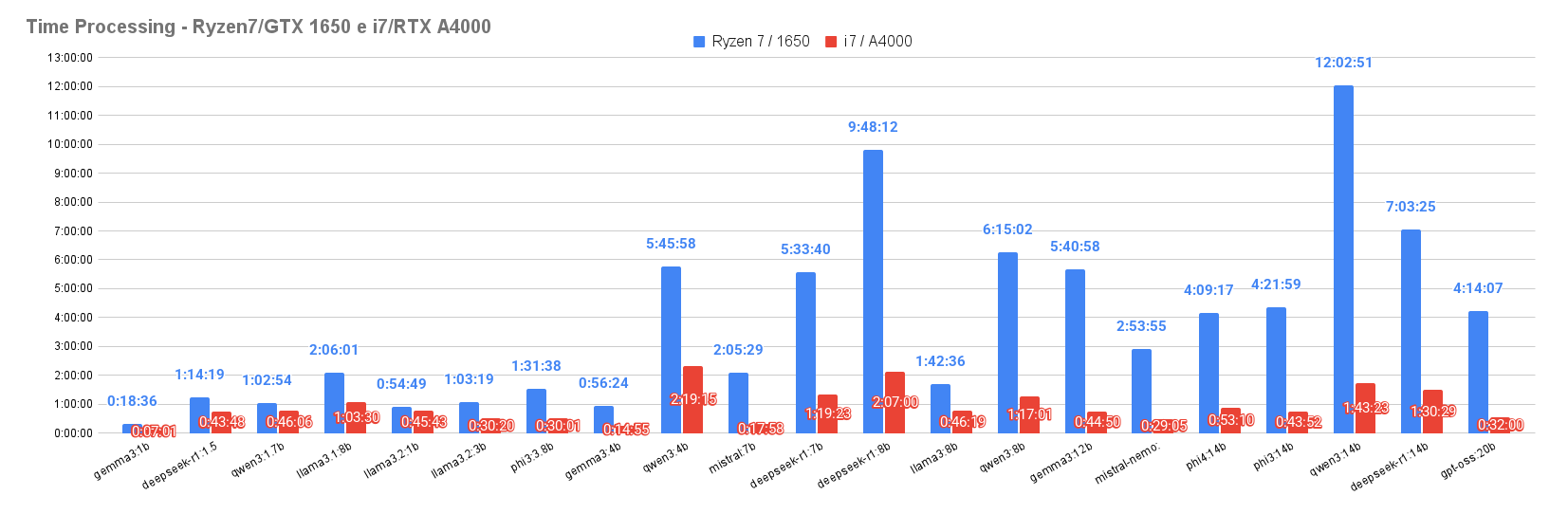}
    \caption{Processing Time per Architecture}
    \label{fig:tempoprocessamento}
\end{figure}

\autoref{fig:tempoprocessamento} shows that the Ryzen7/GTX 1650 architecture exhibited significantly higher execution times, whereas the i7/RTX A4000 architecture achieved substantially faster performance. This difference reflects the greater efficiency of CPU and GPU resource management in the second architecture, highlighting the direct impact of the execution environment on model performance.

As observed, the variation in the execution time between the different temperatures (T0, T0.4, T0.7, and T1) is relatively small, suggesting that the temperature hyperparameter primarily influences the diversity and textual coherence of the responses rather than inference time. However, smaller models such as \textit{gemma3:1b}, \textit{deepseek-r1:1.5b}, and \textit{llama3:1.8b} exhibit execution times far below those of larger models such as \textit{gpt-oss:20b} and \textit{deepseek-r1:14b}. These results reinforce the notion that the number of parameters is one of the decisive factors for computational cost.

In the i7/RTX A4000 architecture, the greater dispersion in execution times can be explained by the way Ollama manages model loading and inference, performing \textit{load/unload} operations and establishing varying communication patterns between CPU and GPU. This dynamic generates different latencies, depending on the model size and the usage of GPU memory.

Smaller models tend to underutilize CUDA cores, whereas larger models require memory reallocation, resulting in fewer linear execution times. The absence of specific optimizations, such as \textit{mixed precision} and tuning for Tensor Core utilization, also contributes to this oscillation. In contrast, the Ryzen7/GTX 1650 architecture exhibited more regular behavior, possibly due to more stable execution and simpler resource management.

\begin{figure}[htbp]
    \centering
    \includegraphics[width=1\textwidth]{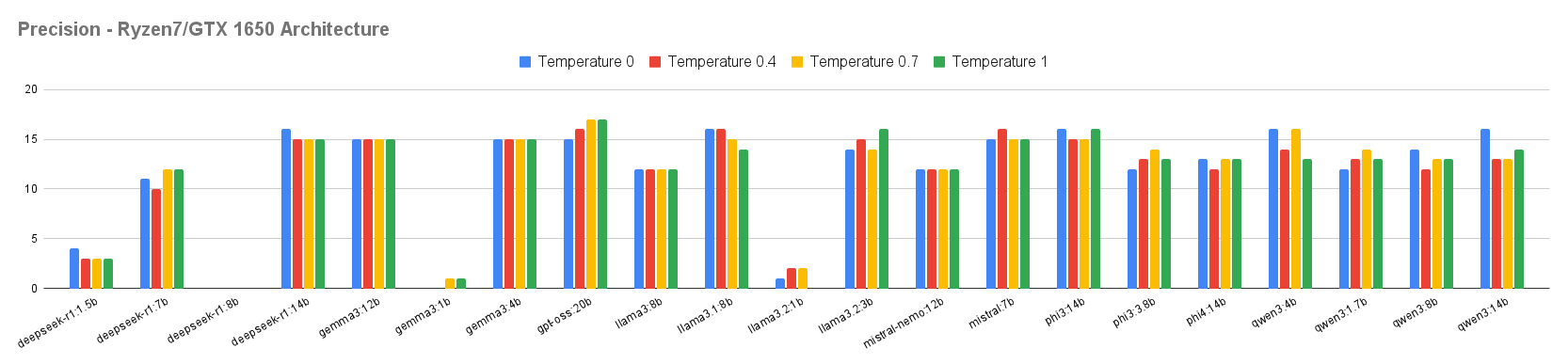}
    
    \vspace{0.3em}
    
    \includegraphics[width=1\textwidth]{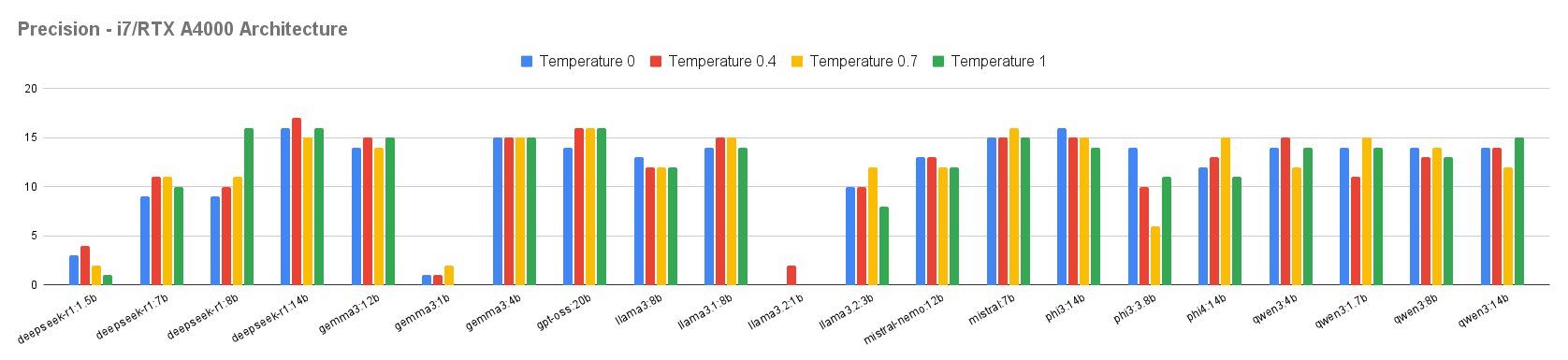}
    \caption{Precision per Architecture}
    \label{fig:diferentes}
\end{figure}

\autoref{tab:precision} and  \autoref{fig:diferentes} present the summary of the average precision of the models. As shown, precision remained relatively stable across the different inference temperatures (T0 to T1), indicating that the temperature hyperparameter exerts limited influence on the precision of classification .

\begin{table}[htbp]
    \centering
    \caption{Model Precision at Different Temperatures}
    \label{tab:precision}
    \setlength{\tabcolsep}{1.5pt} 
    \renewcommand{\arraystretch}{1.3} 
    \begin{tabular}{l|cc|cc|cc|cc|cc|cc|cc|cc}
    \hline
    & \multicolumn{8}{c|}{\textbf{Ryzen7/GTX 1650 Architecture}} & \multicolumn{8}{c}{\textbf{i7/RTX A4000 Architecture}} \\
    \cline{2-17}
    \textbf{Model} & \textbf{T0} & \textbf{\%} & \textbf{T0.4} & \textbf{\%} & \textbf{T0.7} & \textbf{\%} & \textbf{T1} & \textbf{\%} & \textbf{T0} & \textbf{\%} & \textbf{T0.4} & \textbf{\%} & \textbf{T0.7} & \textbf{\%} & \textbf{T1} & \textbf{\%} \\
    \hline
    deepseek-r1:1.5b & 4 & 16,67 & 3 & 12,50 & 3 & 12,50 & 3 & 12,50 & 3 & 12,50 & 4 & 16,67 & 2 & 8,33 & 1 & 4,17 \\
    deepseek-r1:7b & 11 & 45,83 & 10 & 41,67 & 12 & 50,00 & 12 & 50,00 & 9 & 37,50 & 11 & 45,83 & 11 & 45,83 & 10 & 41,67 \\
    deepseek-r1:8b & 0 & 0,00 & 0 & 0,00 & 0 & 0,00 & 0 & 0,00 & 9 & 37,50 & 10 & 41,67 & 11 & 45,83 & 16 & 66,67 \\
    deepseek-r1:14b & 16 & 66,67 & 15 & 62,50 & 15 & 62,50 & 15 & 62,50 & 16 & 66,67 & 17 & 70,83 & 15 & 62,50 & 16 & 66,67 \\
    gemma3:12b & 15 & 62,50 & 15 & 62,50 & 15 & 62,50 & 15 & 62,50 & 14 & 58,33 & 15 & 62,50 & 14 & 58,33 & 15 & 62,50 \\
    gemma3:1b & 0 & 0,00 & 0 & 0,00 & 1 & 4,17 & 1 & 4,17 & 1 & 4,17 & 1 & 4,17 & 2 & 8,33 & 0 & 0,00 \\
    gemma3:4b & 15 & 62,50 & 15 & 62,50 & 15 & 62,50 & 15 & 62,50 & 15 & 62,50 & 15 & 62,50 & 15 & 62,50 & 15 & 62,50 \\
    gpt-oss:20b & 15 & 62,50 & 16 & 66,67 & 17 & 70,83 & 17 & 70,83 & 14 & 58,33 & 16 & 66,67 & 16 & 66,67 & 16 & 66,67 \\
    llama3:8b & 12 & 50,00 & 12 & 50,00 & 12 & 50,00 & 12 & 50,00 & 13 & 54,17 & 12 & 50,00 & 12 & 50,00 & 12 & 50,00 \\
    llama3.1:8b & 16 & 66,67 & 16 & 66,67 & 15 & 62,50 & 14 & 58,33 & 14 & 58,33 & 15 & 62,50 & 15 & 62,50 & 14 & 58,33 \\
    llama3.2:1b & 1 & 4,17 & 2 & 8,33 & 2 & 8,33 & 0 & 0,00 & 0 & 0,00 & 2 & 8,33 & 0 & 0,00 & 0 & 0,00 \\
    llama3.2:3b & 14 & 58,33 & 15 & 62,50 & 14 & 58,33 & 16 & 66,67 & 10 & 41,67 & 10 & 41,67 & 12 & 50,00 & 8 & 33,33 \\
    mistral-nemo:12b & 12 & 50,00 & 12 & 50,00 & 12 & 50,00 & 12 & 50,00 & 13 & 54,17 & 13 & 54,17 & 12 & 50,00 & 12 & 50,00 \\
    mistral:7b & 15 & 62,50 & 16 & 66,67 & 15 & 62,50 & 15 & 62,50 & 15 & 62,50 & 15 & 62,50 & 16 & 66,67 & 15 & 62,50 \\
    phi3:14b & 16 & 66,67 & 15 & 62,50 & 15 & 62,50 & 16 & 66,67 & 16 & 66,67 & 15 & 62,50 & 15 & 62,50 & 14 & 58,33 \\
    phi3:3.8b & 12 & 50,00 & 13 & 54,17 & 14 & 58,33 & 13 & 54,17 & 14 & 58,33 & 10 & 41,67 & 6 & 25,00 & 11 & 45,83 \\
    phi4:14b & 13 & 54,17 & 12 & 50,00 & 13 & 54,17 & 13 & 54,17 & 12 & 50,00 & 13 & 54,17 & 15 & 62,50 & 11 & 45,83 \\
    qwen3:4b & 16 & 66,67 & 14 & 58,33 & 16 & 66,67 & 13 & 54,17 & 14 & 58,33 & 15 & 62,50 & 12 & 50,00 & 14 & 58,33 \\
    qwen3:1.7b & 12 & 50,00 & 13 & 54,17 & 14 & 58,33 & 13 & 54,17 & 14 & 58,33 & 11 & 45,83 & 15 & 62,50 & 14 & 58,33 \\
    qwen3:8b & 14 & 58,33 & 12 & 50,00 & 13 & 54,17 & 13 & 54,17 & 14 & 58,33 & 13 & 54,17 & 14 & 58,33 & 13 & 54,17 \\
    qwen3:14b & 16 & 66,67 & 13 & 54,17 & 13 & 54,17 & 14 & 58,33 & 14 & 58,33 & 14 & 58,33 & 12 & 50,00 & 15 & 62,50 \\
    \hline
    \end{tabular}
\end{table}

Comparing the architectures, a slightly superior performance is observed for the i7/RTX A4000 relative to the Ryzen7/GTX 1650, a result that can be attributed to its higher processing capacity and more advanced set of GPU optimizations.

Medium-sized models such as \textit{deepseek-r1:14b}, \textit{phi3:14b}, and \textit{qwen3:4b} achieved the highest precision rates on both platforms, indicating a good balance between generalization capability and computational cost. Overall, the i7/RTX A4000 architecture also provided greater consistency in the results, especially for models with more parameters.

\section{Conclusion and Future Work}\label{consideracoes}

The experiments showed that the temperature hyperparameter exerts limited influence on accuracy in automated categorization of security incidents and does not significantly affect inference time. In contrast, the computational architecture and the number of parameters proved to be determining factors: smaller models were more efficient in terms of performance, while medium-sized models offered a better balance between cost and accuracy, reinforcing the suitability of SLMs for local (\textit{on-premises}) scenarios with resource constraints.

The DeepSeek-R1 14B model achieved the highest precision on the Ryzen7/GTX 1650 architecture, while GPT-OSS 20B ranked first on the i7/RTX A4000 architecture, demonstrating that the performance of SLMs varies according to the computational resources available. The differences between the architectures were also evident. The i7/RTX A4000 environment exhibited greater variation in execution times, possibly due to how Ollama manages model loading and inference, alternating operations between CPU and GPU. However, the Ryzen7/GTX 1650 system showed more linear behavior, suggesting greater stability and predictability in performance.

For future work, we propose:
(i) measuring CPU and GPU usage in detail in order to correlate resource consumption, latency, and stability;
(ii) investigating lightweight optimization techniques such as quantization, weight pruning, and compression to reduce inference time and memory usage in low-cost architectures; and (iii) analyzing the combined impact of multiple hyperparameters (temperature, \textit{top-k}, \textit{top-p}, and penalties) on the semantic quality of classifications.

\bibliographystyle{ieeetr}
\bibliography{references}

\end{document}